\documentclass[floatfix,aps,prl,showpacs,amsmath,floatfix,preprintnumbers,twocolumn,multirow]{revtex4}
\usepackage{graphicx, subfigure}
\usepackage{color}
\usepackage{natbib}
\usepackage[usenames,dvipsnames]{xcolor}

\newcommand{\ddd}{\displaystyle}
\newcommand{\Mpc}{\,\mathrm{Mpc}}

\def\laq{~\raise 0.4ex\hbox{$<$}\kern -0.8em\lower 0.62
ex\hbox{$\sim$}~}

\def\gaq{~\raise 0.4ex\hbox{$>$}\kern -0.7em\lower 0.62
ex\hbox{$\sim$}~}

\def\beq{\begin{equation}}
\def\eeq{\end{equation}}
\def\bea{\begin{eqnarray}}
\def\eea{\end{eqnarray}}
\def\bean{\begin{eqnarray*}}
\def\eean{\end{eqnarray*}}

   \def\be{\begin{equation}}
   \def\ee{\end{equation}}
   \def\ba{\begin{eqnarray}}
   \def\ea{\end{eqnarray}}

\def\laq{~\raise 0.4ex\hbox{$<$}\kern -0.8em\lower 0.62ex\hbox{$\sim$}~}
\def\gaq{~\raise 0.4ex\hbox{$>$}\kern -0.7em\lower 0.62ex\hbox{$\sim$}~}

\def\beq{\begin{equation}}
\def\eeq{\end{equation}}
\def\bea{\begin{eqnarray}}
\def\eea{\end{eqnarray}}

    \def\be{\begin{equation}}
    \def\ee{\end{equation}}
    \def\ba{\begin{eqnarray}}
    \def\ea{\end{eqnarray}}

\newcommand{\eq}{\begin{equation}}
\newcommand{\eqx}{\end{equation}}
\newcommand{\eqn}{\begin{eqnarray}}
\newcommand{\eqnx}{\end{eqnarray}}

\begin{document}

\preprint{DESY-13-166}

\title{Constraining the primordial power spectrum from SNIa lensing dispersion}

\author{Ido Ben-Dayan$^{1}$, Tigran Kalaydzhyan$^{2}$}

\affiliation{$^1$Deutsches Elektronen-Synchrotron DESY, Theory Group, D-22603 Hamburg, Germany }
\affiliation{$^2$Department of Physics and Astronomy, Stony Brook University, \\
                            Stony Brook, New York 11794-3800, USA}

\begin{abstract}
The (absence of detecting) lensing dispersion of supernovae type Ia (SNIa) can be used as a novel and extremely efficient probe of cosmology.
In this preliminary example we analyze its consequences for the primordial power spectrum.
 The main setback is the knowledge of the power spectrum in the nonlinear regime, $1 \Mpc^{-1}\lesssim k \lesssim 10^2-10^3 \Mpc^{-1}$ up to redshift of about unity. By using the lensing dispersion and conservative estimates in this regime of wave numbers, we show how the current upper bound $\sigma_{\mu}(z\leq1)<0.12$ on \textit{existing data} gives strong indirect constraints on the primordial power spectrum.
The probe extends our handle on the spectrum to a total of $12-15$ inflation e-folds.
These constraints are so strong that they are already ruling out a large portion of the parameter space allowed by PLANCK for running, $\alpha\equiv d n_s/d\ln k$, and running of running, $\beta\equiv d^2 n_s/d \ln k^2$.
The bounds follow a linear relation to a very good accuracy.
A conservative bound disfavors any enhancement above the line $\beta(k_0)=0.036-0.42\,\alpha(k_0)$ and a realistic estimate disfavors any enhancement above the line $\beta(k_0)=0.022-0.44\,\alpha(k_0)$.
\end{abstract}

\vspace {1cm}~


\maketitle
\textbf{Introduction:} Cosmology is becoming a precise science, most notably due to increasing number and quality of measurements.
Utilizing several probes is crucial in breaking degeneracies between cosmological parameters. The combination of CMB, large scale structure (LSS) and Type Ia Supernovae (SNIa) has lead to the emergence of the ``concordance model'' of cosmology.
SNIa are widely used in cosmology due to their small intrinsic dispersion around their mean luminosity.
By observing supernovae at cosmological distances,
we can measure the luminosity-redshift relation $d_L(z)$ and infer cosmological parameters from the mean luminosity.
However, the intrinsic dispersion of SNIa luminosities is
not the only source of scatter in the data. Photons arriving from these `Standard Candles'
are affected by the inhomogeneous matter distribution between the source and observer.
This induces an additional scatter in the luminosity-redshift relation, making it a \textit{stochastic} observable with mean, dispersion, etc.
Therefore, by disentangling this cosmic dispersion from the intrinsic scatter, we can potentially probe background parameters like $\Omega_{m0}$ or fluctuations, i.e. the \textit{power spectrum}.
Our main interest will be the lensing contribution, which dominates the cosmic dispersion at $z\gtrsim0.3$.

We suggest using the lensing dispersion of SNIa as an additional probe of cosmology. The now operational Dark Energy Survey \cite{Bernstein:2011zf} will measure thousands of SNIa, up to redshift $z\sim 1.2$ and LSST  \cite{Abell:2009aa} will measure millions of SNIa. This will reduce statistical errors considerably and increase the chance for detection since the lensing dispersion grows with the redshift at $z\sim1$ \cite{Dodelson:2005zt, Hui:2005nm, BenDayan:2013gc}. With future data, it has been suggested to use the lensing dispersion to constrain certain cosmological properties %the present matter density $\Omega_{m0}$ and $\sigma_8$
  \cite{Hamana:1999rk, Minty:2001jg, Dodelson:2005zt, Amendola:2013twa, Quartin:2013moa}.

In this preliminary paper we analyze the implications of the lensing dispersion $\sigma_\mu(z)$ on the primordial power spectrum. The distance modulus, $\mu=5\log_{10}({d_L(z)}/{\mathrm{10pc}})$,
is a function of the luminosity distance $d_L(z)$ to the source at redshift $z$.
 Existing data analysis has not detected lensing dispersion with enough statistical significance, but has placed an upper bound of $\sigma_{\mu}(z\leq 1) \leq0.12$ for the redshift of up to unity \cite{Jonsson:2010wx} at $95\%$ C.L. and other analyses \cite{Kowalski:2008ez, Conley:2011ku, Kronborg:2010uj} yield similar results. A Bayesian analysis was carried out in \cite{Karpenka:2012ys}, suggesting a very marginal detection of a lensing signal. Finally, in \cite{Betoule:2014frx} the Joint Light-curve Analysis (JLA) compiled $740$ SNIa, reducing considerably systematic errors. For $z\lesssim 1$ The JLA analysis used mean value of \cite{Jonsson:2010wx} $\sigma_{\mu}(z)=0.055z$ and added a ``coherent dispersion'' $\sigma_{coh.}$ to account for any other sources of intrinsic variations. The outcome is $\sigma_{coh.}=0.106\pm0.006$. Moreover, there is a clear trend of $\sigma_{coh.}$ decreasing with redshift. Hence, the JLA analysis gives a model independent estimate of the total observed dispersion.
Given that $\sigma_{total}(z)=\sqrt{\sigma_{\mu}(z)^2+\sigma_{coh.}^2}\leq 0.12$, we find that considering $\sigma_{\mu}(z\leq 1) \leq0.12$ is a rather conservative upper bound, and we shall use this bound in our analysis.

In principle, the primordial power spectrum is not limited to a specific parametrization. In practice,
it is typically parameterized as $P_{k}=A_s(k/k_0)^{n_s(k_0)-1}$, where $k_0$ is a suitable ``pivot scale''. A common, more general form, is when the spectral index $n_s(k)$ is scale dependent, and then expanded around the pivot scale $k_0$,
\begin{align}
n_s(k) = \ddd n_s(k_0)+\ddd\frac{\alpha(k_0)}{2} \ln\frac{k}{k_0}+\ddd\frac{\beta(k_0)}{6} \ln^2\frac{k}{k_0}\,,
\end{align}
where $\alpha$ is typically dubbed the ``running'' of the spectral index, and $\beta$, the ``running of running''. The best constraints on $\alpha,\beta$ with $k_0=0.05 \Mpc^{-1},\,n_s(k_0) \simeq 0.96$ are given by PLANCK \cite{Ade:2013uln} and $\mathrm{Ly_\alpha}$ \cite{Zhao:2012xw}.
 These analyses are only probing the range $H_0\leq k\lesssim 1 \Mpc^{-1}$. The lensing dispersion,
$\sigma_{\mu}$ is sensitive to  $0.01 \lesssim k \lesssim 10^2-10^3 \Mpc^{-1}$, thus giving access to $2-3$ more decades of the spectrum. Hence, $\sigma_{\mu}(z)$ is particularly sensitive to the quasi-linear and non-linear part of the spectrum. In terms of inflation, the direct measurement of $k\lesssim 1\Mpc^{-1}$ corresponds to about $8$ e-folds of inflation, leaving most of the power spectrum of $\sim 60$ e-folds out of reach \footnote{In principle, it is possible that primordial perturbations have been generated only during these $8$ e-folds. However, to solve the homogeneity and horizon problem, one generically requires $\sim15-60$ e-folds. Shutting down the generation of perturbations during the rest of the e-folds is rather tuned.}.  Therefore, even after PLANCK there is still an enormous space of inflationary models allowed. It is therefore of crucial importance to infer as much of the spectrum as possible for a better inflationary model selection. The lensing dispersion constrains additional $4-7$ e-folds, yielding a total of $12-15$ e-folds.

Other methods of probing the primordial spectrum include methods such as $\mathrm{Ly_\alpha}$ \cite{Zhao:2012xw}, the absence of primordial black holes and Ultracompact Minihalos \cite{Alabidi:2013lya, Bringmann:2011ut, Li:2012qha}, measuring spectral distortions of the CMB blackbody spectrum  \cite{Chluba:2012gq, Chluba:2012we, Chluba:2013pya},
  galaxy weak lensing \cite{Miyatake:2013bha}, galaxy correlation functions \cite{Marin:2013bbb} and cluster number counts \cite{Chantavat:2008nu}. All of which have either different systematics, different $k$ range sensitivity, based on future data or some combination of the above.
  Albeit degenerate with other cosmological parameters, $\sigma_{\mu}$ surpasses these methods by actually cutting into the allowed parameter space allowed by PLANCK, using \textit{existing data} only. In a separate publication \cite{Preparation}, we analyze the case, where the power spectrum takes a different ``non-slow-roll'' parametrization such as in cases analyzed in \cite{Chluba:2012we}.

\begin{figure}[t]
\includegraphics[width=7cm]{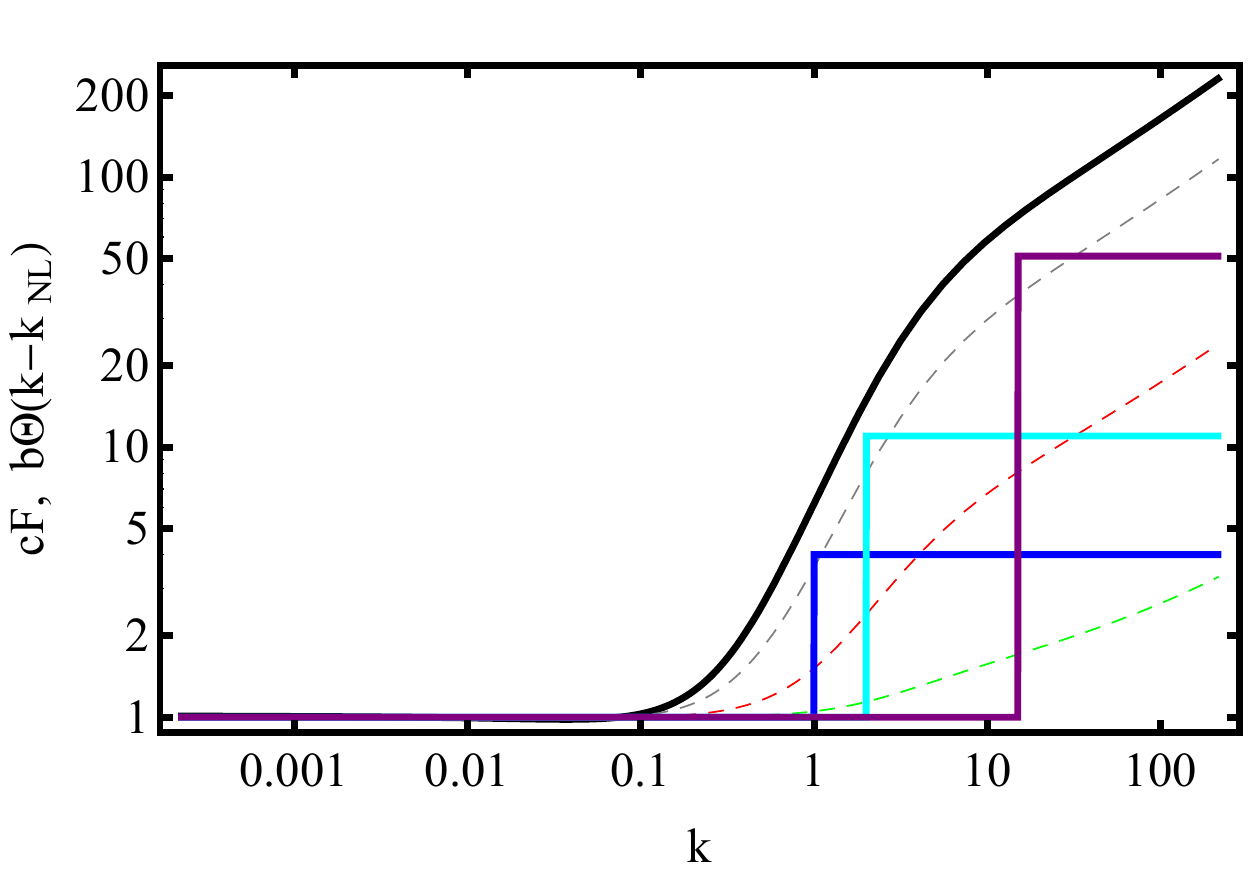}
\caption{Log-Log plot of the ``transfer functions'' (\ref{transfer_function}) at redshift $z=1$ multiplied by $c=1$ (solid black), $c=0.5$ (dashed grey), $c=0.1$ (dashed red) $c=0.01$ (dashed green). Solid blue, cyan and purple curves are the step functions in (\ref{step_funct}) with $b=3, k_{NL}=1\Mpc^{-1}$; $b=10, k_{NL}=2 \Mpc^{-1}$; $b=50, k_{NL}=15 \Mpc^{-1}$, respectively.}\label{TLNL}
\vspace{-0.5cm}
\end{figure}

\textbf{Method:} We start from the full dispersion expression of the luminosity distance, calculated in the light-cone average approach up to second order in the Poisson (longitudinal) gauge, \cite{Gasperini:2011us, BenDayan:2012pp, BenDayan:2012ct, BenDayan:2012wi,  BenDayan:2013gc}, and recently confirmed in \cite{Fanizza:2013doa}. The dominant contribution of the dispersion at $z\gtrsim0.3$, comes from the lensing contribution. For a perturbed FLRW Universe, one starts with the line of sight (LOS) first order lensing contribution to the distance modulus,
\begin{align}
\delta \mu_1(\eta_s^{(0)}) =\frac{5}{\ln 10} \int_{\eta_s^{(0)}}^{\eta_o} \frac{d\eta_1}{\Delta\eta}\frac{\eta_1 - \eta_s^{(0)}}{\eta_o - \eta_1}\Delta_2\Psi
% |_{r_i=\eta_o-\eta_i}
 \,,
 \label{delmu}
\end{align}
where the gravitational potential $\Psi=\Psi(\eta_i,r_i,\tilde\theta^a)$ is evaluated along the past light-cone at $r_i=\eta_o-\eta_i$, $\eta_o$ is the observer conformal time, $\eta_s^{(0)}$ is the conformal time of the source with unperturbed geometry, $\Delta \eta(z)=\eta_o(z)-\eta_s^{(0)}(z)=\int^z_0
\frac{dy}{H_0\sqrt{\Omega_{m0}(1+y)^3+\Omega_{\Lambda0}}}$ and $\Delta_2$ is the 2D angular Laplacian, see \cite{BenDayan:2013gc, BenDayan:2012pp} for technical terms and explanations. Squaring (\ref{delmu}) and taking the ensemble average in Fourier space at a fixed observed redshift, gives the variance, $\sigma^2_{\mu}(z)$ \footnote{In \cite{BenDayan:2012pp, BenDayan:2012ct, BenDayan:2013gc} we have used a combination of light cone and ensemble average, which meant an integral over the sky 2-sphere to derive a similar result. It is unnecessary and the ensemble average here is sufficient. This makes our analysis useful even in the case of sparse data. We thank an anonymous referee and Dominik Schwarz for comments that lead to this conclusion.}. In general, the calculation involves a complicated double line of sight (LOS) and wave number integration. However, at $z\gtrsim0.3$, the double LOS integral is dominated by the equal time part, and further by $\mathrm{Si}(x \gg 1) \approx \pi/2$, yielding:
\begin{align}
\label{sigmuLNL}
 \sigma_{\mu}^2 &\simeq \left(\frac{5}{ \ln 10}\right)^2 \frac{\pi}{\Delta\eta^2} \int_{\eta_s^{(0)}}^{\eta_o} \frac{d\eta_1 dk}{k}P_{\Psi}(k,\eta_1)k^3\\
 & ~~~~~~~~~~~~~~~~~~~~~~~~~~~~\times (\eta_1-\eta_s^{(0)})^2(\eta_o-\eta_1)^2\,,\nonumber
\end{align}
where $P_{\Psi}$ is the linear (LPS, $P_L$) or non-linear dimensionless power spectrum (NLPS, $P_{NL}$) of the \textit{gravitational potential} \footnote{A similar formula can be derived for any perturbed FLRW cosmology in which photons fulfill the geodesic equations of general relativity, such as dark energy models. The only change is the time dependent behavior of the gravitational potential $\Psi$.}.
Hence, the lensing dispersion of supernovae is a direct measurement of the integrated late-time power spectrum. At the most basic level, this can be used to constrain parameterizations of $P_{\Psi}$, or cosmological parameters such as $\Omega_{m0},\sigma_8$ or $w(z)$.
 We will be mostly interested in the dispersion at $z=1$ where sufficient data is available and because up to the redshift of a few, the dispersion grows approximately linearly \cite{Holz:2004xx, Kronborg:2010uj, Jonsson:2010wx, BenDayan:2013gc}, so the best constraints can be given at the maximal available redshift.
 In general, the $k^2$ enhancement makes $\sigma_{\mu}$ a sensitive probe to the small scales of the power spectrum. To make this claim transparent, let us switch to dimensionless variables, $\tilde \eta= H_0 \eta$ and $p=k/k_{eq}$ \footnote{The choice of the equality scale $p=k/k_{eq}$ is because we know the general behaviour of $P_L$, or more precisely, its transfer function $T(k)$ which is constant for $p<1$ and scales like $p^{-2} \ln p$ for $p \gg 1$}:
 \begin{align}
\label{dimless}
 \sigma_{\mu}^2 \simeq \left(\frac{5}{ \ln 10}\right)^2 \frac{\pi}{\Delta \tilde \eta^2} \left(\frac{k_{eq}}{H_0}\right)^3 \int d\tilde \eta_1 dp P_{\Psi}(p,\tilde \eta_1)p^2\\
\times (\tilde \eta_1-\tilde \eta_s^{(0)})^2(\tilde \eta_o-\tilde \eta_1)^2\,.\nonumber
 \end{align}
From this expression we learn that: $(a)$ The relevant physical scales are $H_0$ and $k_{eq}$, which give an enhancement of $(k_{eq}/H_0)^3$. $(b)$ The dispersion is really sensitive to the scales smaller than the equality scale $p>1$. $(c)$ The NLPS has an additional redshift dependent physical scale which is the onset of nonlinearity. For a given redshift, parameterizing the NLPS as $\sim C (k/k_{NL})^{\nu}$, from some $k_{NL}$, will have an additional parametric enhancement of $(k_{NL}/k_{eq})^3$.
\begin{figure*}
\subfigure{\includegraphics[width=7cm]{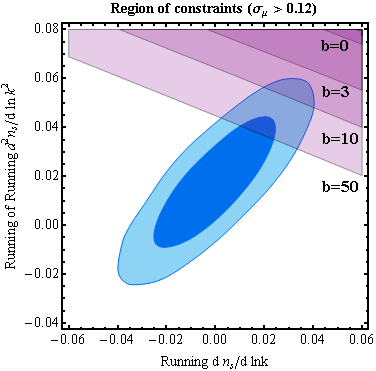}}\hspace{1cm}
\subfigure{\includegraphics[width=7cm]{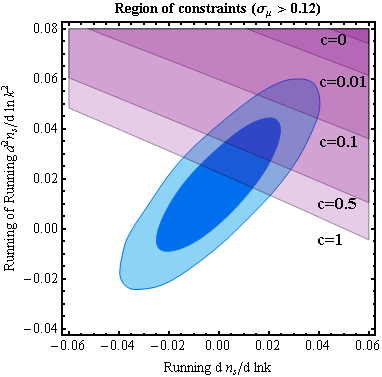}}
\vspace{-0.3cm}
\caption{Regions of allowed parameters combined with  PLANCK data. The ellipses are the 68\% and 95\% C.L. contours from \cite{Ade:2013uln}. In the colored regions $\sigma_{\mu}(z=1)>0.12$ and are disfavored for $b=0,1,3,10,50$ (left panel) and $c=0,0.01,0.1,0.5,1$ (right panel)}\label{planck_constr}
\end{figure*}

 Equation (\ref{dimless}) is free of IR and UV divergences so the main limitation is the validity of the spectrum \cite{BenDayan:2013gc}. For $k\gg H_0$ at some redshift dependent point standard cosmological perturbation theory breaks down, and one has to resort to numerical simulations to get an approximate fitting formula for the power spectrum. We use the HaloFit model \cite{Smith:2002dz, Takahashi:2012em, Inoue:2012px} with $k_{UV}=320h \Mpc^{-1}$. We have verified that varying $k_{UV}\in(30h,\infty) \Mpc^{-1}$ or $H_0$ and $\Omega_{m0}$ independently within the range $H_0\in[64, 70]$, $\Omega_{m0}\in[0.27, 0.36]$ gives at most $15\%$ change in the value of $\sigma_{\mu}$. Taken that $\sigma_{\mu}(z=1, \alpha=\beta=0, k_{UV}=320h \Mpc^{-1})=0.08$, the bound cannot be saturated by varying the background parameters and/or integrating up to arbitrarily small scales. Hence the bound can be useful for constraining $\alpha(k_0)$ and $\beta(k_0)$ and our results are accurate to about $20\%$.

Considering a big enough $\alpha$ and $\beta$, the HaloFit fitting formula is not reliable anymore due to its sensitivity to initial conditions. For example, with $\alpha=0.04$ and $\beta=0.05$ the existing HaloFit actually gives enhancement of a few at $1<k<175 \,\Mpc^{-1}$ and a \textit{suppressed} power spectrum compared to the linear one at larger $k$. It is nevertheless obvious that the non-linear evolution causes clustering and enhances the power spectrum. For example, at redshift $z=1$, the ratio between the HaloFit formula, $P_{NL}(k,z)$, with standard initial conditions $(n_s\simeq0.96,\, \alpha=\beta=0)$  and the linear power spectrum $P_{NL}(k,1)/P_L(k,1)$ is the solid, thick, black curve plotted in Fig.~\ref{TLNL}. Already at $k=1 \Mpc^{-1}$ the non-linear power spectrum is a factor of a few larger than the linear one, and for $k\gtrsim 10 \Mpc^{-1}$, it behaves as a power law with a scaling exponent of nearly $1/2$.
We therefore utilize this ratio in the standard case of $n_s=const.$ to define a ``transfer function'',
\bea
F(k,z)&\equiv &\frac{P_{NL}(k,z)}{P_L(k,z)}\,, \label{transfer_function}
\eea
where $P_{NL}$ is the non-linear power spectrum, $P_L=(3/5)^2 P_k T^2(k)g^2(z)$ is the linear spectrum and $T(k)$ is the transfer function with baryons \cite{Eisenstein:1997ik}, all taken in the standard scenario with $n_s\simeq0.96,\alpha=\beta=0$. We take the enhancement into account in two simple ways. The first method is by the Heaviside function $\Theta (k)$. Here we are not limited to the HaloFit formula, so we perform the following substitution in equation (\ref{sigmuLNL}),
\be
P_{\Psi} \rightarrow P_L(k,z)(1+b\,\Theta(k-k_{NL}))\label{step_funct}
\ee
and we evaluate $\sigma_{\mu}$ for $b=0,3,10,50$ with corresponding $k_{NL}=1,1,2,15 \Mpc^{-1}$, such that the step function is always underestimating the transfer function $F$, so this is a very conservative estimate.
The step functions are the solid blue, cyan and purple lines in Fig.~\ref{TLNL}.
The second method is to use $F$ of the HaloFit model, such that
\be
P_{\Psi}\rightarrow P_L(k,z)(1-c+c F(k,z))\,,
\ee
and evaluate $\sigma_{\mu}$ with $c=0,0.01,0.1,0.5,1$. In both methods $b=0$ or $c=0$ correspond to computing the dispersion with the linear power spectrum only, while
$c=1$ corresponds to exactly following the HaloFit enhancement pattern. Except $c=1$ all the second method values of $c$ are underestimates as well. The resulting enhancement at $z=1$, is plotted in Fig.~\ref{TLNL} as green, red and grey dashed lines.

\textbf{Results:} In Fig.~\ref{planck_constr} we show the constraints on running and running of running from the non-detection of lensing dispersion overlaid on PLANCK likelihood contours. In the left panel, the values $b=0,3,10,50$ with corresponding $k_{NL}=1,1,2,15 \Mpc^{-1}$ are considered. The right panel considers $c=0,0.01,0.1,0.5,1$. In both panels, colored regions give $\sigma_{\mu}(z=1)\geq0.12$ and are disfavored.

We wish to note that there are additional factors which make our analysis an underestimate.  First of all,
partial sky coverage is expected to increase dispersion \cite{Hui:2005nm}. Second, SNIa at higher redshift have already been detected and used for cosmological parameter inference. The monotonicity of $\sigma_{\mu}(z)$ ensures that considering, for instance, $\sigma_{\mu}(z=1.2)$ would give more stringent bounds. Third, the consideration of other analyses.
Bayesian analyses \cite{March:2011xa,Karpenka:2012ys} suggested that the total dispersion is about $0.12$ with a very marginal detection of the lensing signal. Better yet, the JLA \cite{Betoule:2014frx} is an up to date, model independent analysis and also there $\sigma_{total}(z\leq1)\leq 0.12$, not just the lensing dispersion. In the above cases, the intrinsic or ``coherent'' dispersion, actually dominates the total dispersion. On top of that, in the JLA analysis there is a clear trend of $\sigma_{coh.}$ decreasing in redshift, meaning that the actual value of the lensing dispersion is probably smaller than the $\sigma_{\mu}=0.055z$ it uses. Last, all other analyses (data, statistical, theoretical and numerical) \cite{Kronborg:2010uj, Jonsson:2010wx, BenDayan:2013gc,Holz:2004xx,Smith:2013bha} point to a lower value of the dispersion as well, at most $\sigma_{\mu}(z)\simeq0.093z$, practically disfavouring even a larger portion of the parameter space allowed by PLANCK.

\textbf{Conclusions and Outlook:} From Fig.~\ref{planck_constr}, it is obvious that the lensing dispersion or its absence is an extremely powerful cosmological probe. Even if a scale dependent spectral index induces clustering which is \textit{an order of magnitude smaller} than the standard constant $n_s$ scenario, some of the parameter space allowed by PLANCK is ruled out. Moreover, the analysis discusses the spectrum up to $k\sim320 h\Mpc^{-1}$, more than two orders of magnitude beyond PLANCK's lever arm ($\sim5$ e-folds more) irrespective of whether models are ruled in or out. It can be treated as a prediction of inflationary models. In the more realistic case where the enhancement is similar to the HaloFit model, such as $c=0.5,1$,  one gets strong bounds on the allowed parameters, that can be expressed as a linear relation,
\bea
\beta(k_0) \leq0.036-0.42\,\alpha(k_0), \quad c=0.5\\
\beta(k_0)\leq0.022-0.44\,\alpha(k_0), \quad c=1.
\eea
 The realistic case of $\beta(k_0)\leq 0.022$
nicely matches PLANCK's $\alpha(k_0)=0^{+0.016}_{-0.013}\,,\beta(k_0)=0.017_{-0.014}^{+0.016}$.
 Obviously, a definite detection of lensing will enable a more stringent analysis similar to CMB lensing.

It is very appealing to add the lensing dispersion constraint to the likelihood analysis of the PLANCK data. We believe that numerical simulations with initial conditions as suggested here, $\alpha(k_0),\beta(k_0)\neq0$, which will give a more accurate late time power spectrum, will yield similar results, thus strengthening our argument. These simulations are already on their way.
  Last, we have suggested using the (absence of) dispersion to constrain the primordial power spectrum. Since the dispersion depends on several cosmological parameters, it can be useful in constraining other fundamental cosmological parameters as well.

\textbf{Acknowledgements}
It is a pleasure to thank Matthias Bartelmann, Torsten Bringmann, Jens Chluba, Thomas Konstandin, Alexander Westphal and Mathias Zaldarriaga for helpful discussions, comments and suggestions.
The work of I.B.-D. is supported by the German Science Foundation (DFG) within the
Collaborative Research Center (CRC) 676 Particles, Strings and the Early Universe. The work of T.K. was supported in part by the U.S. Department of Energy under Contract DE-FG-88ER40388.

\bibliography{draft10}

\end{document}